\documentclass[english,keywords,aps,amsmath,amssymb,twocolumn]{revtex4-1}
\usepackage[T1]{fontenc}
\usepackage[latin1]{inputenc}
\usepackage{babel}
\usepackage{amssymb}
\usepackage{graphicx}
\usepackage{color}
\usepackage{bm}
\usepackage{longtable}
\usepackage{amsmath} 
\usepackage{amsfonts}
\usepackage{amssymb}
\newcommand{\be}{\begin{equation}}
\newcommand{\ee}{\end{equation}}
\newcommand{\ba}{\begin{eqnarray}}
\newcommand{\ea}{\end{eqnarray}}

\begin{document}
\title{Device-independent certification of Hilbert space dimension using a family of Bell expressions}
\author{ A. K. Pan }
\email{akp@nitp.ac.in}
\author{Shyam Sundar Mahato}
\affiliation{National Institute of Technology Patna, Ashok Rajpath, Patna, Bihar 800005, India}
\begin{abstract}

Dimension witness provides a device-independent certification of the minimal dimension required to reproduce the observed data without imposing assumptions on the functioning of the devices used to generate the experimental statistics. In this paper, we provide a family of Bell expressions where Alice and Bob perform $2^{n-1}$ and $n$ number of dichotomic measurements respectively which serve as the device-independent dimension witnesses of Hilbert space of $2^{m}$ dimensions with $m=1,2..2^{\lfloor n/2\rfloor}$.  The family of Bell expressions considered here determines the success probability of a communication game known as $n$-bit parity-oblivious random access code.  The parity obliviousness constraint is equivalent to  preparation non-contextuality assumption in an ontological model of an operational theory. For any given $n\geq 3$, if such a constraint is imposed on the encoding scheme of the random-access code, then the local bound of the Bell expression reduces to the preparation non-contextual bound.  We provide explicit examples for $n=4,5$ case to demonstrate that the relevant Bell expressions certify the qubit and two-qubit system, and for $n=6$ case, the relevant Bell expression certifies the qubit, two-qubit and three-qubit systems. We further demonstrate the sharing of quantum preparation contextuality by multiple Bobs sequentially to examine whether number of Bobs sharing the preparation contextuality is dependent on the dimension of the system. We provide explicit example of $n=5$ and $6$ to demonstrate that number of Bobs sequentially sharing the contextuality remains same for any of the $2^{m}$ dimensional systems. 
\end{abstract}
\maketitle
\section{Introduction}
Dimensionality is a fundamental property of a quantum system.  The Hilbert space in which the quantum state belongs is an abstract construction but the number of dimension available to a system is a physical quantity and is considered to be resource for quantum computation and quantum information theory \cite{robin}. Higher dimensional system can make a given protocol more efficient and alternatively, the security of many cryptographic protocols relies on the dimensional characteristics of the system. For example, instead of qubit system if  four-dimensional states are used, then the celebrated Bennett-Brassard (BB84) cryptographic protocol \cite{bb84} can be shown to be entirely compromised \cite{acin06,brunner11}. From the fundamental perspective, there are quantum correlations whose simulation by classical resources inevitably require dimensional superiority. The quantum dimension witness is a criterion that provides a lower bound on the dimension that is needed to reproduce a given measurement statistics. Of late, the device-independent dimension witness has become an important research area where the dimension of a quantum system is certified without requiring \emph{a priori} knowledge about the devices used in the experiment. 

The notion of dimension witness was first introduced in a seminal paper by Brunner \textit {et al.,}{\tiny }\cite{brunner} in the context of bipartite Bell scenario, which involve two spatially separated observers Alice and Bob,  who access uncharacterized devices (black boxs). Alice and Bob receive inputs $x\in\{1,2,...n_{A}\}$ and $y\in\{1,2.....n_{B}\}$ respectively and the uncharacterized measurement device yielding respective outputs $a\in\{0,1\}$ and $b\in\{0,1\}$. The conditional probability $P(ab|xy)$ admits a $d$-dimensional representation if it can be written as;
$P(ab|xy) = tr[\rho_{AB} (M_{a}^{x}\otimes M_{b}^{y})]$ for the state $\rho_{AB}$ $\in$ $\mathbb{C}^{d} \otimes\mathbb{C}^{d}$ shared between two parties and the local measurements $M_{a}^{x}$ and $M_{b}^{y}$ acting on $\mathbb{C}^{d}$. The reproduction of every joint probability $P(ab|xy)$ in quantum theory puts a lower bound on the dimension of the Hilbert space. 

Since then flurry of interesting works along this direction have been reported \cite{wehner, gallego, dal12,brunnerprl13,guh14,bowler,bowlers,muk15, sik16,sik16prl,cai16,cong17,vin17,son17,cze18,str19,vin19,zha19,spee20}. The work of Brunner \textit {et al.,} \cite{brunner} was further generalized and extended to prepare measure scenario by Gallego \emph{et al.,} \cite{gallego} and proposed a family of inequalities which serve as classical and quantum dimension witnesses are given by 
\begin{equation}
\mathbf{I}_{N}=\sum_{y=1}^{N-1} E_{1y} + \sum_{x=2}^{N} \sum_{y=1}^{N+1-x} \alpha_{xy} E_{xy}
\end{equation}
where, $\alpha_{xy} =1 $ if $ x + y \leq N$ , and $\alpha_{xy} =-1 $ otherwise. Here, $x\in \{1,2,....N\}$, $y\in \{1,2.....N-1\}$ and $E_{xy}$ is the correlation. The problem of dimension witness is meaningful if the  number of preparation $(N)$ is greater than the Hilbert space dimension of the system. For classical states of dimension $d\le N$ it is found that algebraic bound $I_{N}\le L_{d}$, where $L_{d} = \dfrac{N(N-3)}{2} + 2d -1$. For example $N=3$ and $d=2$ one finds the classical value is 3 and quantum value is $2\sqrt{2}+1$. Further analysis found that $I_{3}$ inequity is achieve its optimal value for $d=3$ which is $5$. So this inequality has the ability to test the dimension as well as to distinguish between the classical and quantum system.

Later, by assuming independence of the prepare and measure devices a non-linear dimension witness is also proposed \cite{bowler} and test of dimension in communication network is also proposed \cite{bowlers}. A connection to random access code \cite{wehner} and to state discrimination \cite{brunnerprl13} are also pointed out.  In a recent proposal, by employing binary-outcome measurements a certification of an arbitrary-dimensional quantum systems is proposed \cite{cze18}.  Experimental verifications of  dimension witness including higher dimensional system has also been performed \cite{ahrens,hend,ahrens2,Ambrosio,zhu16,sun16,agulier,guo20}.  

In this paper we provide a family of quantum dimension witnesses based on the parity-oblivious random access code \cite{spek09}. The parity oblivious condition  imposed on Alice's encoding scheme implies here that no parity information of the inputs of Alice is shared to Bob. It can be shown \cite{ghorai18} that the success probability of a $n$-bit random access code can be solely determined by a family of Bell expressions $(\mathcal{B}_{n})$ where Alice and Bob use $2^{n-1}$ and $n$ number of dichotomic measurements respectively. Importantly, for a given $n$, the optimal quantum value of $\mathcal{B}_{n}$ can only be achieved for the quantum system having local Hilbert space dimension $d=2^{\lfloor n/2\rfloor}$.  The parity oblivious constraint is shown \cite{spekk05} to be equivalent to  preparation non-contextuality assumption in an ontological model, and for a given $n$, such a constraint on the encoding scheme  reduce the local bound of the family of Bell expressions $\mathcal{B}_{n}$ to preparation non-contextual bound \cite{spekk05}. This is due to the fact that such a condition puts further restriction on free choices of the values of Alice's observables.

For $n=2$ and $n=3$, the Bell expressions are well-known CHSH \cite{chsh} and elegant Bell expressions \cite{gisin} respectively. Since both the Bell expressions for $n=2,3$ can be optimized for qubit system, they cannot be served as dimension witness of Hilbert space. However, each of the Bell expressions for $n\geq 4$ has the potential to distinguish  the dimensions $d=2^{m}$ of the Hilbert space with $m=1,2..2^{\lfloor n/2\rfloor}$ thereby serve as dimension witness of the Hilbert space.  We provide explicit examples for $n=4,5$ and $6$ cases to demonstrate that the Bell expressions for both $n=4$ and $n=5$ certify qubit and two-qubit systems and for $n=6$ the relevant Bell expression certifies the qubit, two-qubit and three-qubit local systems. 

Further, we examine the sharing of preparation contextuality by multiple sequential Bobs perform unsharp measurements. Using the family of Bell expressions mentioned above, it was shown \cite{asmita19} that the sharing of preparation contextuality can be demonstrated for arbitrary number of Bobs by using optimal quantum value of the family of Bell expressions, achieved for the $d=2^{\lfloor n/2\rfloor}$ dimensional Hilbert space. We argue that that there is a possibility of sharing preparation contextuality by arbitrary number of Bobs for the system in lower dimensional Hilbert space. We provide explicit example for $n=5$ where the number of sequential Bobs sharing preparation contextuality remains same for qubit and two-qubit system. Similarly, for $n=6$ the sharing is for possible for same number of sequential Bobs for qubit, two-qubit and three qubit systems. However, the value of unshrapness parameter required for demonstrating preparation contextuality by sequential Bob is always higher in lower dimensional system, as expected. 

This paper is organized as follows. In Sec. II, we briefly recapitulate the essence of parity-oblivious random-access-code and derivation of the family of preparation non-contextual inequalities, i.e., the Bell inequalities those serve as the dimension witnesses. In Sec.III, we provide the sum-of-square approach to optimize the dimension witnesses for various dimensional quantum systems. We provide the specific examples for $n=4,5$ and $6$ to demonstrate how the corresponding Bell inequalities certify the $2^{m}$ dimensional systems in Sec. IV. For the dimension witnesses for $n=4,5$ and $6$, we examine the sharing of preparation contextuality by multiple Bobs for qubit, two-qubit and three-qubit systems in Sec. V. Finally, we summarize our work in Sec. VI.  

\section{A family of Bell expressions serve as dimension witnesses}
Since the family of dimension witnesses is based on the parity-oblivious random access code (PORAC) and the parity-oblivious condition is equivalent to preparation non-contextuality assumption in an ontological model of an operational theory, we first provide the essence of preparation non-contextuality and then, the derivation of local and preparation non-contextual bound of the aforementioned family of Bell expressions. 

We start by encapsulating the notion of an ontological model reproducing the quantum statistics \cite{hari,spekk05}. In quantum theory, a preparation procedure $(P)$ produces a density matrix $\rho$ and the measurement procedure $(M)$ which is in general described by a suitable positive-operator-valued-measure (POVM) $E_k$, provides the probability of occurrence an outcome $ k $ is given by $p(k|P, M)=Tr[\rho E_{k}]$, which is the Born rule. In an ontological model of quantum theory, the preparation of quantum state $\rho$ by a specific preparation procedure $P$  is equivalent to preparing a probability distribution $\mu_{P}(\lambda|\rho)$ in the ontic state space, satisfying $\int _\Lambda \mu_{P}(\lambda|\rho)d\lambda=1$ where $\lambda \in \Lambda$ and $\Lambda$ is the ontic state space. The probability of obtaining an outcome $k$ is a response function $\xi_{M}(k|\lambda, E_{k}) $ satisfying $\sum_{k}\xi_{M}(k|\lambda, E_{k})=1$ where a measurement operator $E_{k}$ is realized through a  measurement procedure $M$. The primary requirement of such an ontological model is to reproduce the Born rule, i.e., $\forall \rho $, $\forall E_{k}$ and $\forall k$, $\int _\Lambda \mu_{P}(\lambda|\rho) \xi_{M}(k|\lambda, E_{k}) d\lambda =Tr[\rho E_{k}]$.

The notion of non-contextuality was reformulated and generalized for any operational theory by Spekkens\cite{spekk05}. For our purpose, we focus on the quantum theory here.  An ontological model of quantum theory can be considered to be preparation non-contextual if  $\forall M :  p(k|P, M)=p(k|P^{\prime}, M)\Rightarrow \mu_{P}(\lambda|\rho)=\mu_{P^{\prime}}(\lambda|\rho)$ is satisfied where $P$ and $P^{\prime}$ are two distinct preparation procedures but in the same equivalent class \cite{kunjwal, mazurek,hameedi,pan19}. 

As mentioned, the family of dimension witnesses in the present work are derived through a two-party communication game known PORAC. It was shown by Spekkens \cite{spek09} that the parity-oblivious condition in an operational theory can equivalently be cast into the assumption of preparation non-contextuality in an ontological model. It is already shown in \cite{ghorai18} that the success probability of a $n$-bit PORAC can be solely linked with a family of Bell's inequities. For the sake of completeness, we briefly encapsulate encapsulate the essence of the derivation. 

In a  $n$-bit PORAC, Alice chooses her bit $x^{\delta}$ randomly from $\{0,1\}^{n}$ with $\delta\in \{1,2...2^{n}\}$. The relevant ordered set $\mathcal{D}_n$ can be written as  $\mathcal{D}_n = (x^{\delta} | x^i \oplus x^j = 111...11 \;\text{and}\; i+j = 2^n +1)$ and $i \in \{1,2, ...2^{n-1}\}$. Here, $ x^1 = 00...00, x^2 = 00...01, .... $, and so on. Bob can choose any bit $ y \in \{1,2, ..., n\}$ and recover the bit $x^{\delta}_y$ with a probability. The condition of the task is the following;  Bob's output must be the bit $b=x^{\delta}_y$. The parity-oblivious constraint here is that \emph{no} information about any parity of $ x $ can be transmitted to Bob. Following \cite{spek09}, we define a parity set $ \mathbb{P}_n= \{x|x \in \{0,1\}^n,\sum_{r} x_{r} \geq 2\} $ with $r\in \{1,2,...,n\}$. For any arbitrary $s \in \mathbb{P}_{n}$, no information about $s.x = \oplus_{r} s_{r}x_{r}$ (s-parity) is to be transmitted to Bob, where $\oplus$ is sum modulo $ 2 $.

%Our problem is estimated from the parity oblivious multiplexing task (POM). In \cite{ghorai} showed that the quantum success probability provided in \cite{spekken} is the optimal value for the 3-bit POM task. They first prove that the optimal quantum success probability for 3-bit POM task is solely dependent on the optimal quantum violation of the elegant Bell inequality proposed by Gisin \cite{gisin}. They also generalized it upto n-bit POM task which contain $2^{n-1}$ and $n$ measurement on the Alice and Bob side respectively. We identify their generalized inequality and relate it to dimension witness. We have shown that there is no appearance of dimension witness upto $n=3$. Bell expression that we have used that give dimension witness when n>3.we derive the Bell expression in Eq.(\ref{nbell}) using parity- oblivious Random access code as a tool.\linebreak 

In an operational theory, Alice encodes her $ n $-bit string of $ x^{\delta} $ prepared by a procedure $P_{x^{\delta}}$. Next, after receiving the particle, for every $y \in \{1,2,...,n\}$, Bob performs a two-outcome measurement $M_{n,y}$ and reports outcome $b$ as his output. Then the probability of success is given by
\begin{equation}
\label{qprob}
p(b=x^{\delta}_{y}) = \dfrac{1}{2^n n}\sum\limits_{x,y}p(b=x^{\delta}_y|P_{x^{\delta}},M_{n,y}).
\end{equation}

%Let Alice and Bob share an entangled state $\rho_{AB}=|\psi_{AB}\rangle \langle\psi_{AB}| \in\mathbb{C}^{d}\otimes \mathbb{C}^{d}$ and the goal is to estimate the minimal dimension of the shared entangled state from the experimental statistics $P(ab|xy)$ where $a$ and $b$ are the measurement outcomes $a,b\in \{-1,1\}$. 

 \begin{figure}[ht]
 \centering 
{{\includegraphics[width=0.7\linewidth]{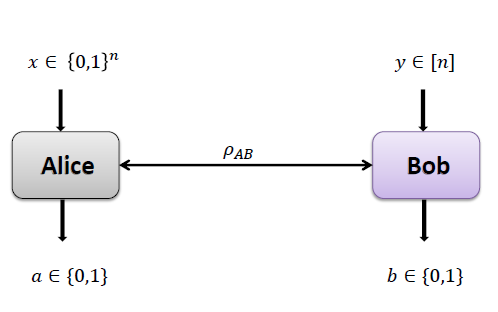}}}
\caption{Schematic diagram of bipartite Bell scenario}
 \end{figure}
In quantum PORAC, Alice encodes her $n$-bit string of $ x^{\delta} $ into quantum states $ \rho_{x^{\delta}} $, prepared by a procedure $P_{x^{\delta}}$. On a suitable entangled state $ \rho_{AB}$ = $|{\psi_{AB}}\rangle  \langle{\psi_{AB}}|$ with $|\psi_{AB}\rangle \in \mathbb{C}^{d}\otimes\mathbb{C}^{d}$,  Alice performs one of the $2^{n-1}$ projective measurements $\{{P_{A_{n,i}}}$, $\mathbb{I}-P_{A_{n,i}}\}$ where $ i \in \{1, 2, ... 2^{n-1}\}$ to encode her $n$-bits into $ 2^n $ quantum states are given by-
\begin{eqnarray}
\label{states}
\nonumber
\dfrac{1}{2} \rho_{x^i}& =& Tr_A[(P_{A_{n,i} }\otimes \mathbb{I}) \rho_{AB}]\\
\dfrac{1}{2} \rho_{x^j}& =& Tr_A[(\mathbb{I}-P_{A_{n,i}} \otimes \mathbb{I}) \rho_{AB}] 
\end{eqnarray}
with $ i+j=2^n+1 $. 

In quantum theory, the parity-oblivious condition implies that 
\begin{align}
\label{poc}
	\forall s: \frac{1}{2^{n-1}}\sum\limits_{x^{\delta}|x^{\delta}.s=0} \rho_{x^{\delta}}=\frac{1}{2^{n-1}}\sum\limits_{x^{\delta}|x^{\delta}.s=1} \rho_{x^{\delta}}
\end{align}

In ontological model of quantum theory, the parity-obliviousness in Eq.(\ref{poc})  condition is equivalent to the preparation non-contextual assumption, i.e., 

\begin{align}
	\forall s: \frac{1}{2^{n-1}}\sum\limits_{x^{\delta}|x^{\delta}.s=0} \mu(\lambda|\rho_{x^{\delta}})=\frac{1}{2^{n-1}}\sum\limits_{x^{\delta}|x^{\delta}.s=1} \mu(\lambda|\rho_{x^{\delta}})
\end{align}

Note that the number of parity-oblivious conditions for $n$-bit PORAC is the number of element in $ \mathbb{P}_n$ \cite{ghorai18}. We noticed that there are two types of parity oblivious conditions. The one arising from the natural construction, such as, $\mathbb{I}=P_{A_{n,i}}^{+}+P_{A_{n,i}}^{-}$. In that case, $s \in \mathbb{P}_n $ follow the property $ \sum_y s_y = 2m$ with $m \in \mathbb{N} $.  For the rest of $s \in \mathbb{P}_n$ not satisfying the above property, a non-trivial constraints on Alice's observables need to be satisfied are given by

\begin{equation}
\label{oba}
\sum_{i=1}^{2^{n-1}} (-1)^{s.x^i} A_{n,i} = 0
\end{equation}
The total number of such non-trivial constraints on Alice's observables is  $C_n= 2^{n-1} - n$.

The measurements of for decoding scheme are taken to be 
\begin{eqnarray}
M_{n,y} &=& \begin{cases}
M_{n,y}^i,\text{when}\; b=x_y^i \\ M_{n,y}^j, \text{when}\; b=x_y^j
\end{cases};\\
M_{n,y}^{i(j)} & =& \begin{cases}
P_{B_{n,y}},& \text{when}\; x^{i(j)}_y=0 \ \\ \mathbb{I} - P_{B_{n,y}},& \text{when}\; x^{i(j)}_y=1 
\end{cases}	
%M_{y}^{j} = \begin{cases}
%P_{B_y}, & \text{when}\; x^j_y=0 \\ \mathbb{I} - P_{B_y}, & \text{when}\; x^j_y=1
%\end{cases}	
\end{eqnarray}
The quantum success probability can then be written as 
\begin{eqnarray}
\label{succ1}
\nonumber
p_{Q} & =&  \dfrac{1}{2^n n} \sum_{y=1}^{n} \sum\limits_{i=1}^{2^{n-1}} p(b=x^i_y|\rho_{x^i},M_{y}^i) + p(b=x^j_y|\rho_{x^j},M_{y}^j)\\
 &=&\dfrac{1}{2} + \dfrac{	\mathcal{B}_{n} }{2^n n}
\end{eqnarray} 
where $\mathcal{B}_{n}$ is the family of Bell expressions is given by
\begin{align}
\label{nbell}
\mathcal{B}_{n} =  \sum_{y=1}^{n}\sum_{i=1}^{2^{n-1}} (-1)^{x^i_y}  A_{n,i}\otimes B_{n,y}
\end{align}
which serve as the family of dimension witnesses in the present work. Note that, the Bell expressions in Eq.(\ref{nbell}) provides the CHSH and Gisin's elegant Bell's inequality for $n=2$ and $n=3$ respectively and the corresponding local bounds are 2 and 6. The local bound of the family of Bell expressions for any arbitrary $n$ is given by \cite{ghorai18}
\begin{equation}
\label{lbound}
(\mathcal{B}_{n})_{local}\leq n {n-1 \choose \lfloor \dfrac{n-1}{2} \rfloor}
\end{equation}

However, the parity-oblivious condition in the usual RAC imposes a functional relationship between Alice's observables as given by Eq. (\ref{oba}). This means that Alice choices of values are restricted and the local bound of Eq. (\ref{nbell}) gets reduced (which we call preparation-noncontextual bound \cite{spek09,ghorai18}) to 
\ba
\label{pncbound}
(\mathcal{B}_{n})_{pnc}\leq 2^{n-1}
\ea
Since for any $n$, we have $(\mathcal{B}_{n})_{pnc}\leq (\mathcal{B}_{n})_{local}$ then for a given $n$, even if optimal quantum value $(\mathcal{B}_{n})_{Q}$  does not violate the local bound in Eq. (\ref{lbound}), it may still reveal non-classicality by violating the preparation non-contextual given by Eq. (\ref{pncbound}).  Note that, it is already known \cite{ghorai18} that the optimal quantum value of $\mathcal{B}_{n}$ can be obtained for the local system having dimension $d=2^{\lfloor n/2\rfloor}$. Our purpose here to examine the maximum quantum values of $\mathcal{B}_{n}$ that can be achieved for the lower dimensional systems having dimensions $d<2^{\lfloor n/2\rfloor}$.   

\section{Sum-of-square approach for maximization}
In order to find the quantum upper bound of the Bell expression $(\mathcal{B}_{n})_{d=2^{m}}$ for various dimensions,  we use sum-of-square (SOS) approach (see, for example, \cite{bamps}), so that $(\mathcal{B}_{n})_{Q}\leq \beta_{n}$  for all possible quantum states $\rho_{AB}$ and measurement operators $A_{n,i}$ and $B_{n,y}$. Here  $\beta_{n}$ is the upper bound on the quantum  value of $(\mathcal{B}_{n})_{d=2^{m}}$ for the system having dimension $d=2^{m}$. This is equivalent to showing that there is a positive semidefinite operator $\gamma_{n}\geq 0$, that can be expressed as $\langle \gamma_{n}\rangle_{Q}=\beta_{n}-(\mathcal{B}_{n})_{Q} $where $\beta_{n}$ is a number. This can be proven by considering a set of suitable positive operators $M^{i}_{n}$ which is polynomial functions of   $A_{n,i}$ and $B_{n,y}$, so that 
\begin{align}
\label{gamma}
		\gamma_{n}=\sum\limits_{i=1}^{2^{n-1}}  \frac{\omega_{n,i}}{2}(M^{i}_{n})^{\dagger}M^{i}_{n}
\end{align}
 where $\omega_{n,i}$ is positive semidefinite and to be specified shortly. The maximum value of $(\mathcal{B}_{n})_{Q}$  for a any given dimension is obtained if $\langle \gamma_{n}\rangle_{Q}=0$, implying that 

\begin{align}
\label{mm}
	M^{i}_{n}|\psi\rangle=0
\end{align}

For the family of Bell expressions given by Eq. (\ref{nbell}), the operators $M^{i}_{n}$ can be written as

\begin{align}
\label{mi}
	M^{i}_{n}=\frac{1}{\omega_{n, i}}\sum\limits_{y=1}^{n} (-1)^{x^i_y} B_{n,y} -A_{n,i}
\end{align}
where $	\omega_{n,i}=||\sum\limits_{y=1}^{n} (-1)^{x^i_y} B_{n,y}||$.  Plugging Eq. (\ref{mi}) into Eq. (\ref{gamma}) and by noting that $A_{n,i}^{\dagger} A_{n,i}=B_{n,y}^{\dagger} B_{n,y}=\mathbb{I} $, we get
\begin{align}
\langle \gamma_{n}\rangle_{Q}=-(\mathcal{B}_{n})_{Q} + \sum\limits_{i=1}^{2^{n-1}}\left[ \frac{1}{2\omega_{n,i}}\left(\sum\limits_{y=1}^{n} (-1)^{x^i_y} B_{n,y}\right)^2 +  \frac{\omega_{n,i}}{2}\right]
\end{align}
which can be re-written in a simple form as

\begin{align}
\langle \gamma_{n}\rangle_{Q}=-(\mathcal{B}_{n})_{Q} + \sum\limits_{i=1}^{2^{n-1}}\omega_{n,i}
\end{align}
Maximum quantum value of $(\mathcal{B}_{n})_{Q}$ can be obtained when $\langle\gamma_{n}\rangle_{Q}= 0$ which in turn provides 

\begin{align}
\label{optbn}
(\mathcal{B}_{n})_{Q}^{max} &= \underset{B_{n,y}}{max}\left(\sum\limits_{i=1}^{2^{n-1}}\omega_{n,i}\right)\\
\nonumber
&=\underset{B_{n,y}}{max}\left(\sum\limits_{i=1}^{2^{n-1}}||\sum\limits_{y=1}^{n} (-1)^{x^i_y} B_{n,y}||\right)
\end{align}
and explicit condition obtained from Eq. (\ref{mm}) is given by

\begin{align}
\label{conda}
	\forall i \ \ \ \sum\limits_{y=1}^{n} (-1)^{x^i_y} B_{n,y} |\psi\rangle =\omega_{n,i} A_{n,i}|\psi\rangle
\end{align}
To obtain the maximum quantum value from Eq. (\ref{optbn}) for a given dimensional system, we use the concavity inequality, i.e., 
\begin{align}
\label{concav}
	\sum\limits_{i=1}^{2^{n-1}}\omega_{n,i}\leq \sqrt{2^{n-1} \sum\limits_{i=1}^{2^{n-1}} (\omega_{n,i})^{2}}
\end{align}
In Eq. (\ref{concav}), the equality can be obtained only when  $\omega_{n,i}$ are equal for each $i$, when $B_{n,y}$ are mutually anti-commuting and thus for $n>3$ optimal value cannot be obtained for qubit system.

Also, for satisfying the parity-obliviousness  conditions the Eq. (\ref{oba}) has to be satisfied by Alice's observables $A_{n,i}$.  This implies that the optimal quantum value of Bell expression $(\mathcal{B}_{n})$ can only be achieved for bounded dimension of the Hilbert space. By using Eqs. (\ref{oba}) and (\ref{conda}) the condition that is required to hold is given by

\begin{align}
\label{cond2}
	\sum\limits_{i=1}^{2^{n-1}}\sum\limits_{y=1}^{n} (-1)^{s.x^{i}+x^i_y} \frac{B_{n,y}}{\omega_{n,i} }=0
	\end{align}
Since $B_{n,y}$s are dichotomic, the quantity $\omega_{n,i}$ can explicitly be written as 

\begin{align}
\label{omega}
	\omega_{n,i}&= \Big[ n+ \{(-1)^{x^i_1} B_{n,1}, \sum\limits_{y=2}^{n}(-1)^{x^i_y} B_{n,y}\} \\
	\nonumber
	&+\{(-1)^{x^i_2} B_{n,2}, \sum\limits_{y=3}^{n}(-1)^{x^i_y} B_{n,y}\} +........ \\
	\nonumber
	&+\{(-1)^{x^i_{n-1}} B_{n,n-1}, (-1)^{x^i_n} B_{n,n}\}\Big]^{-1/2}
\end{align}
where $\{,\}$ denotes the anti-commutation.

As already mentioned that the optimal quantum value $(\mathcal{B}_{n})_{Q}^{opt}$  of the family of Bell expressions can only be achieved when  Bob's observables $B_{n,y}$ are mutually anti-commuting and this in turn fixes the dimension of the Hilbert space $d=2^{\lfloor n/2\rfloor}$. In such a case $\omega_{n,i}=\sqrt{n}$ for every $i$ and from Eq. (\ref{optbn}) the optimal quantum value can be calculated as 

\begin{align}
\label{opt}
(\mathcal{B}_{n})_{d=2^{\lfloor n/2\rfloor}}^{opt}= 2^{n-1}\sqrt{n}
\end{align}
For this, required maximally entangled state having local  dimension  $ 2^{\lfloor n/2\rfloor} $  is given by

\[|\phi\rangle_{AB} = \dfrac{1}{\sqrt{2^{\lfloor n/2\rfloor}}} \sum\limits_{k=1}^{2^{\lfloor n/2\rfloor}}  |k\rangle_A |k\rangle_B \].

 Thus, for the cases $n\geq 4$, the qubit system will not suffice the purpose and one requires higher dimensional system. For example, the optimal value of the Bell expression for $n=4$ requires at least two-qubit system and for qubit system an upper bound $(\mathcal{B}_{4})_{d=2}^{max}$ can be found, which is smaller than the optimal quantum value. Hence, $\mathcal{B}_{4}$ serves as a dimension witness for certifying the qubit system. Similarly, for any arbitrary $n\geq4$, the Bell expression $\mathcal{B}_{n}$ given by Eq. (\ref{nbell}) certifies  $d=2^{m}$ dimensional local system with $m=1,2,.., \lfloor n/2 \rfloor$. 

\section{Dimension witnesses for one-, two- and three-qubit systems}
In the following, we provide several examples starting from the case of $n=3$ to $n=6$. As already discussed, optimal value of Bell expressions $\mathcal{B}_{n}$ for $n=2$ and $n=3$ require two and three mutually anti-commuting observables respectively,  which can be obtained for qubit system and thus $\mathcal{B}_{2}$ and $\mathcal{B}_{3}$ cannot serve as dimension witness. But from $n\geq 4$ the Bell expression $\mathcal{B}_{n}$ serves as dimension witness. We explicitly demonstrate that the Bell expressions $\mathcal{B}_{4}$ and $\mathcal{B}_{5}$ for $n=4$ and $n=5$ respectively, serve as dimension witnesses for qubit and two qubit systems, and the Bell expression for $n=6$ serves as dimension witness for qubit, two-qubit and three-qubit systems. We first provide the analysis for $n=3$ to make the reader familiar with the optimization technique and how parity-oblivious conditions are satisfied by Alice's observables when optimal quantum value is achieved for qubit system.   
\subsection{Analysis for $n=3$}

For $n=3$,  from Eq. (\ref{nbell}) we obtain Gisin's elegant Bell expression \cite{gisin} is given by
\begin{eqnarray}
\label{bell3}
\nonumber
\mathcal{B}_{3}& =&A_{3,1} \otimes \left(B_{3,1}+B_{3,2}+B_{3,3}\right) \\
\nonumber
&+& A_{3,2} \otimes \left(B_{3,1}+B_{3,2}-B_{3,3}\right)\\
\nonumber
&+&A_{3,3} \otimes \left(B_{3,1}-B_{3,2}+B_{3,3}\right) \\
\nonumber
&+& A_{3,4} \otimes \left(-B_{3,1}+B_{3,2}+B_{3,3}\right)
\end{eqnarray}
The local bound of $\mathcal{B}_{3}$ is $6$. The parity-oblivious condition derived from Eq. (\ref{oba}) provides a functional relation between the Alice's observables, i.e., $A_{3,1}-A_{3,2}-A_{3,3}-A_{3,4}=0$.  If this condition is imposed, the local bound reduces to the preparation non-contextual bound $4$. The optimal quantum value of the relevant Bell expression is $(\mathcal{B}_{3})_{Q}^{opt} = max\left(\sum\limits_{i=1}^{4}\omega_{3,i}\right) $ where $\omega_{3,i}$ can be written as 
\ba
\nonumber
\omega_{3,1}&=\sqrt{3+ \{B_{3,1},\left(B_{3,2} +B_{3,3}\right)\}+\{B_{3,2}, B_{3,3}\}}\\
\omega_{3,2}&=\sqrt{3+ \{B_{3,1},\left(B_{3,2} -B_{3,3}\right)-\{B_{3,2}, B_{3,3}\}}\\
\nonumber
\omega_{3,3}&=\sqrt{3+ \{B_{3,1},\left(B_{3,2} -B_{3,3}\right)\}-\{B_{3,2}, B_{3,3}\}}\\
\nonumber
\omega_{3,4}&=\sqrt{3- \{B_{3,1},\left(B_{3,2} +B_{3,3}\right)\}+\{B_{3,2}, B_{3,3}\}}
\ea
 For $n=3$, by noting a symmetry, we have $\sum\limits_{i=1}^{4}(\omega_{3,i})^{2}=12$ which can only be available if $B_{3,1}$, $B_{3,2}$ and $B_{3,3}$   are mutually commuting. This in turn provides $\omega_{3,i}=\sqrt{3}$ for each $i$ and thereby providing  $(\mathcal{B}_{3})_{d=2}^{opt} = 4\sqrt 3$. 

From Eq. (\ref{conda}), one can find the observables  $A_{3,i}$ required  for Alice to obtain the optimal violation of the elegant Bell inequality.  Such a choice can be available for qubit system by taking mutually anti-commuting observables of Bob, viz., $B_{3,1} = \sigma_{x}$ ,$B_{3,2} = \sigma_{y}$  and $B_{3,3} = \sigma_{z}$. Using Eq.(\ref{conda}), Alice's choices of observables are the following; $A_{3,1} = (\sigma_{x} + \sigma_{y}+\sigma_{z})/{\sqrt{3}}$, $A_{3,2}=(\sigma_{x} + \sigma_{y} - \sigma_{z})/{\sqrt{3}}$, $A_{3,3}= (\sigma_{x} - \sigma_{y} + \sigma_{z})/{\sqrt{3}}$, $A_{3,4}=(-\sigma_{x} + \sigma_{y} + \sigma_{z})/{\sqrt{3}}$.   

Such choices of observables by  Alice need to satisfy the parity oblivious condition given by Eq. (\ref{cond2}). Using Eqs. (\ref{conda}) and (\ref{cond2}) we find that the following conditions has to be satisfied by $\omega_{3,i}=1/{\alpha_{3,i}}$ are given by

\ba
\label{ocond}
\nonumber
\alpha_{3,1}-\alpha_{3,2}-\alpha_{3,3}+\alpha_{3,4}=0\\
\alpha_{3,1}-\alpha_{3,2}+\alpha_{3,3}-\alpha_{3,4}=0\\
\nonumber
\alpha_{3,1}+\alpha_{3,2}-\alpha_{3,3}-\alpha_{3,4}=0
\ea
The solutions of the Eq. (\ref{ocond}) are $\omega_{3,1}=\omega_{3,2}=\omega_{3,3}=\omega_{3,4} \equiv \omega_{3}^{\prime}$. This is only possible if $B_{3,y}$ are mutually anti-commuting and in this case  $\omega_{3}^{\prime}=\sqrt{3}$, as expected. As mentioned, for $n=3$ the relevant Bell expression can be optimized for qubit system and hence does not serve as dimension witness. However, we demonstrate below that for $n\geq 4$ the family of Bell expressions serve as the dimension witnesses of the Hilbert space.

\subsection{Dimension witness for $n=4$}
Next, we demonstrate that for $n\geq4$ the Bell expressions Eq. (\ref{nbell}) serve as  witnesses  of the Hilbert space having dimension $d=2^{m}$ where $m=1,2... \lfloor n/2\rfloor$. We first demonstrate that for $n=4$ the maximum quantum value of Bell expression for qubit system is smaller than the optimal value obtained for two-qubit system, i.e.,  $(\mathcal{B}_{4})_{d=2}^{max}\leq(\mathcal{B}_{4})_{d=2^{2}}^{opt}$. 

The Bell expression for $n=4$ can be written as
\begin{eqnarray}
\label{bell4}
\nonumber
\mathcal{B}_{4}&=A_{4,1} \otimes\left( B_{4,1}+B_{4,2}+B_{4,3}+B_{4,4}\right)\\
\nonumber
&+A_{4,2} \otimes\left( B_{4,1}+B_{4,2}+B_{4,3}-B_{4,4}\right)\\
\nonumber
&+A_{4,3} \otimes\left( B_{4,1}+B_{4,2}-B_{4,3}+B_{4,4}\right)\\
\nonumber
&+A_{4,4} \otimes\left( B_{4,1}-B_{4,2}+B_{4,3}+B_{4,4}\right)\\
&+ A_{4,5} \otimes\left( -B_{4,1}+B_{4,2}+B_{4,3}+B_{4,4}\right)\\
\nonumber
&+A_{4,6} \otimes\left( B_{4,1}+B_{4,2}-B_{4,3}-B_{4,4}\right)\\
\nonumber
&+A_{4,7} \otimes\left( B_{4,1}-B_{4,2}+B_{4,3}-B_{4,4}\right)\\
\nonumber
&+A_{4,8} \otimes\left( B_{4,1}-B_{4,2}-B_{4,3}+B_{4,4}\right)
\end{eqnarray}

whose local bound is $12$ and preparation non-contextual bound is $8$. As already mentioned, by using the concavity inequality Eq. (\ref{concav})  one finds the optimal value $(\mathcal{B}_{3})_{2^{2}}^{opt}=16$ for two-qubit system when all the $\omega_{4,i}$ are equal to $2$. This happens when all four $B_{4,y}$ are mutually anti-commuting in two-qubit system. One such choice is $ B_{4,1} = \sigma_{x} \otimes \sigma_{x}, B_{4,2} = \sigma_{x} \otimes \sigma_{y},  B_{4,3}= \sigma_{x} \otimes \sigma_{z}$ and $B_{4,4} = \sigma_{y} \otimes \mathbb{I}$, and three more such sets are possible. However, for qubit system there are only three mutually commuting observables are available and then $\mathcal{B}_{3}$ cannot reach the optimal value for qubit system.

We derive the maximum quantum value of $(\mathcal{B}_{4})_{d=2}^{max}$ for qubit system. It is straightforward to understand that all of the eight $\omega_{4,i}$  from Eq. (\ref{optbn}) cannot be equal for qubit system. The question is that how many of them are equal to each other. Using the concavity inequality in Eq. (\ref{concav}) two times one finds that there are two optimal sets for which at most four of them are equal to each other. For example, $\omega_{4,1}=\omega_{4,4}=\omega_{4,5}=\omega_{4,6}\equiv \omega_{4}^{\prime}$ and   $\omega_{4,2}=\omega_{4,3}=\omega_{4,7}=\omega_{4,8}\equiv \omega_{4}^{\prime\prime}$. This provides  the condition on Bob's observables are given by  
\ba
\label{bb1}
&&\{B_{4,1}, B_{4,2}\}=\{B_{4,1}, B_{4,3}\}=\{B_{4,1}, B_{4,4}\}=0\\
\nonumber
&&\{B_{4,2}, B_{4,3}\}=\{B_{4,2}, B_{4,4}\}=0
\ea 

From Eq. (\ref{optbn}), we can then write 
\ba
\label{rac4}
(\mathcal{B}_{4})_{Q}&\leq&4\left(\omega_{4}^{\prime}+\omega_{4}^{\prime\prime}\right)\\
\nonumber
&=&4\left( \sqrt{4+\{B_{4,3}, B_{4,4}\}}+\sqrt{4-\{B_{4,3}, B_{4,4}\}}\right)
\ea
It is easy to check from Eq. (\ref{rac4}) that the optimal value $(\mathcal{B}_{4})_{Q}^{opt}$ can be obtained only when $\{B_{4,3}, B_{4,4}\}=0$ along with the relations in Eq. (\ref{bb1}), i.e., all four $B_{4,y}$ are mutually anti-commuting. Such a requirement cannot be fulfilled for a qubit system and one needs at least two-qubit system.

 Now, for a qubit system, it can be checked that $\mathcal{B}_{4}$ reaches its maximum value 

\ba
(\mathcal{B}_{4})_{d=2}^{max} = 4\left(\sqrt{2}+\sqrt{6}\right)\leq (\mathcal{B}_{4})_{d=2^{2}}^{opt}=16
\ea
when $\{B_{4,3}, B_{4,4}\}=\pm1$.  Explicitly, the choices of $B_{4,y}$ are $B_{4,1}=\sigma_{x}$, $B_{4,1}=\sigma_{y}$, $B_{4,1}=\sigma_{z}$ and $B_{4,1}=\pm \sigma_{z}$. Thus, the Bell expression $\mathcal{B}_{4}$ is a dimension witness distinguishing the dimension between qubit and two-qubit Hilbert space. In the Appendix A, we demonstrate how the Alice's choices observables required to obtain the maximum quantum values of $\mathcal{B}_{4}$ for qubit and two-qubit systems satisfy the parity-oblivious condition.

\subsection{Dimension witness for $n=5$}

We now demonstrate that the Bell expression in Eq. (\ref{nbell}) for $n=5$ also serves as a dimension witness for qubit system.  The explicit form of $\mathcal{B}_{5}$ is given in Eq. (\ref{b5}) of the Appendix B. Following the same technique adopted for $n=3$ and $n=4$ we can find that for optimizing $\mathcal{B}_{5}$ the following relations between  $B_{5,y}$ has to be satisfied; $\{B_{5,1},B_{5,2}\}=\{B_{5,2},B_{5,3}\}=\{B_{5,1},B_{5,3}\}=\{B_{5,3},B_{5,4}\}=\{B_{5,3},B_{5,5}\}=\{B_{5,4},B_{5,5}\}=0$, $\{B_{5,1},B_{5,4}\}=\{B_{5,1},B_{5,5}\}$ and $\{B_{5,2},B_{5,4}\}=-\{B_{5,2},B_{5,5}\}$. Using those relations, from Eq. (\ref{optbn}), we find 
\ba
\label{rac5}
\nonumber
(\mathcal{B}_{5})_{Q}&&\leq 4 \sqrt{5+ \{\left(B_{5,1}+B_{5,2}\right),\left(B_{5,4}+B_{5,5}\right)\}}\\
&+& 4\sqrt{5+\{\left(B_{5,1}-B_{5,2}\right),\left(B_{5,4}-B_{5,5}\right)\}}
\ea

Note that, the optimal quantum value can be reached if $\{B_{5,1},B_{5,4}\}=\{B_{5,1},B_{5,5}\}=\{B_{5,2},B_{5,4}\}=\{B_{5,2},B_{5,5}\}=0 $ which means all $B_{5,y}$ are mutually anti-commuting observables providing $(\mathcal{B}_{5})_{d=2^{2}}^{opt}=16\sqrt{5}$. Again, such choices cannot be obtained for a qubit system and one requires at least two-qubit system. A choice of such set of observables are given by $ B_{5,1} = \sigma_{x} \otimes \sigma_{x}, B_{5,2} = \sigma_{x} \otimes \sigma_{y}, B_{5,3}= \sigma_{x} \otimes \sigma_{z}, B_{5,4} = \sigma_{y} \otimes \mathbb{I}$ and $B_{5,5} = \sigma_{z} \otimes \mathbb{I}$. 

Now, for qubit system the maximum quantum value can be obtained for the following choices of the observables; $B_{5,1} = \sigma_{x}, B_{5,2} = \sigma_{y}, B_{5,3} = \sigma_{y}, B_{5,4}= \left(\sigma_{x} + \sigma_{z}\right)/\sqrt{2}$ and $ B_{5,5} = \left(\sigma_{x} - \sigma_{z}\right)/\sqrt{2}$ and the maximum quantum value is
\ba
(\mathcal{B}_{5})_{d=2}^{max} &=& 8\left(\sqrt{5+2\sqrt{2}}+\sqrt{5-2\sqrt{2}}\right)\\
\nonumber&\leq& (\mathcal{B}_{5})_{d=2^{2}}^{opt}=16\sqrt{5}
\ea
Alice's observables can be found by using Eq. (\ref{conda}). In order to examine whether Alice's observables satisfy the parity oblivious conditions when the maximum quantum values for qubit and two-qubit systems, we follow the similar procedures adopted for the cases $n=3$ and $n=4$. The details of the argument is placed in Appendix B.  
\subsection{Dimension witness (for $n=6$)}
We have just demonstrated that the Bell inequalities for $n=4$ and $n=5$  certify the qubit and two-qubit systems. Next, we demonstrate that the Bell expression $\mathcal{B}_{6}$ for $n=6$ certifies qubit, two-qubit and three qubit systems. The explicit form of $\mathcal{B}_{6}$ is quite lengthy but can easily be obtained from Eq. (\ref{nbell}). Once again, to obtain the optimal quantum value of $\left(\mathcal{B}_{6}\right)_{Q}^{opt}=32\sqrt{6}$ one requires all the $32$ values of $\omega_{6,i}$ in Eq. (\ref{omega}) are equal. This can only be obtained if all six $B_{6,y}$ are mutually anti-commuting and thus requires at least three-qubit system. For qubit and two-qubit system we can obtain two upper bounds. Using the concavity relation eight times, we find that the following relations between $B_{6,y}$ has to be satisfied;   $\{B_{6,1}, B_{6,2}\}=\{B_{6,3}, B_{6,4}\}=\{B_{6,5}, B_{6,6}\}=0$ and $\{B_{6,1}, B_{6,3}\}=\{B_{6,1}, B_{6,4}\}=-\{B_{6,2}, B_{6,3}\}-\{B_{6,2}, B_{6,4}\}$. This provides 
\begin{widetext}
\ba
\label{b6qubit}
\nonumber
(\mathcal{B}_{6})_{Q}&\leq&4\left( \sqrt{6+ \{\left(B_{6,1}+ B_{6,2}+B_{6,3}+ B_{6,4}\right), \left(B_{6,5}+ B_{6,6}\right)\}}+ \sqrt{6- \{\left(B_{6,1}+B_{6,2}+B_{6,3}+ B_{6,4}\right), \left(B_{6,5}+ B_{6,6}\right)\}}\right)\\
\nonumber
&+&4\left( \sqrt{6+ \{\left(B_{6,1}- B_{6,2}+B_{6,3}+ B_{6,4}\right), \left(B_{6,5}+ B_{6,6}\right)\}}+ \sqrt{6-\{\left(B_{6,1}-B_{6,2}+B_{6,3}+ B_{6,4}\right), \left(B_{6,5}+ B_{6,6}\right)\}}\right)\\
\nonumber
&+&4\left( \sqrt{6+ \{\left(B_{6,1}+ B_{6,2}+B_{6,3}- B_{6,4}\right), \left(B_{6,5}- B_{6,6}\right)\}}+ \sqrt{6- \{\left(B_{6,1}+ B_{6,2}+B_{6,3}- B_{6,4}\right), \left(B_{6,5}- B_{6,6}\right)\}}\right)\\
\nonumber
&+&4\left( \sqrt{6+ \{\left(B_{6,1}+ B_{6,2}-B_{6,3}+ B_{6,4}\right), \left(B_{6,5}- B_{6,6}\right)\}}+ \sqrt{6- \{\left(B_{6,1}+ B_{6,2}-B_{6,3}+ B_{6,4}\right), \left(B_{6,5}- B_{6,6}\right)\}}\right)\\
\ea
\end{widetext}
The optimal value for the qubit is $(\mathcal{B}_{6})_{d=2}^{max} = 12\sqrt{2}+8\sqrt{6}+12\sqrt{10}$ and choice of observables required are given by
\begin{eqnarray}
 \label{b6qubit}
 \nonumber
 B_{6,1} &=& \frac{\left(\sigma_{y} + \sigma_{x}\right)}{\sqrt{2}}  ; \hspace{0.1cm} B_{6,2} = \frac{\left(\sigma_{y} - \sigma_{x}\right)}{\sqrt{2}}; \hspace{0.1cm} B_{6,3} = \frac{\left(\sigma_{x} + \sigma_{z}\right)}{\sqrt{2}}\\
 \nonumber
 B_{6,4} &=& \frac{\left(\sigma_{x} - \sigma_{z}\right)}{\sqrt{2}}  ; \hspace{0.1cm} B_{6,5} = \frac{\left(\sigma_{z} + \sigma_{y}\right)}{\sqrt{2}}; \hspace{0.1cm} B_{6,6} = \frac{\left(\sigma_{z} - \sigma_{y}\right)}{\sqrt{2}}\\
 \end{eqnarray}

For two-qubit system, we additionally have $\{B_{6,1}, B_{6,5}\}=\{B_{6,2}, B_{6,5}\}=\{B_{6,3}, B_{6,5}\}=\{B_{6,4}, B_{6,5}\}=\{B_{6,2}, B_{6,6}\}=\{B_{6,3}, B_{6,6}\}=\{B_{6,4}, B_{6,6}\}=\{B_{6,5}, B_{6,6}\}=  0$. We can the write Eq. (\ref{b6qubit}) as 
\ba
\label{b62qubit}
\nonumber
(\mathcal{B}_{6})_{d=2^{2}} \leq 16\left( \sqrt{6+ \{B_{6,1}, B_{6,6}\}}+\sqrt{6- \{B_{6,1}, B_{6,6}\}}\right)\\
\ea
In order to get maximum quantum value of $\mathcal{B}_{6}$ for two-qubit system we need to choose $B_{6,y} = B_{5,y} \ \text{for} \  y=1,2..5$ and $\  B_{6,6} = \pm \sigma_{z} \otimes \mathbb{I}$ and the maximum quantum value will be $(\mathcal{B}_{6})_{d=2^{2}}^{max}=32(1+\sqrt{2})$. For a three-qubit system, we additionally have $\{B_{6,1}, B_{6,6}\}=0$, i.e., six mutually anti-commuting observables  available for three-qubit system are given by  
\begin{eqnarray}
 \label{b63qubit}
 &&B_{6,y} = \sigma_{x} \otimes B_{5,y} \ \text{for} \  y=1,2..5 \\
\nonumber
&&\  B_{6,6} = \sigma_{y} \otimes \mathbb{I}\otimes \mathbb{I}
 \end{eqnarray}

We can then summarize the quantum values of $(\mathcal{B}_{6})_{Q}$ for qubit, two-qubit and three-qubit systems as

%\ba
	%&&\{B_{6,1}, B_{6,3}\}=\{B_{6,1}, B_{6,4}\}=-\{B_{6,2}, B_{6,3}\}\\
	%\nonumber
	%&&=-\{B_{6,2}, B_{6,4}\}=\{B_{6,3}, B_{6,5}\}=\{B_{6,3}, B_{6,6}\}\\
	%\nonumber	
	%&&=-\{B_{6,4}, B_{6,5}\}=-\{B_{6,4}, B_{6,6}\}
%\ea

\ba
(\mathcal{B}_{6})_{d=2}^{max} &=& 12\sqrt{2}+8\sqrt{6}+12\sqrt{10}\\
\nonumber
&\leq& (\mathcal{B}_{6})_{d=2^{2}}^{max}=32\Big(1+\sqrt{2}\Big)\\
\nonumber
&\leq& (\mathcal{B}_{6})_{d=2^{3}}^{opt}=32\sqrt{6}
\ea
Hence, the Bell expression $\mathcal{B}_{6}$ can certify the qubit, two-qubit and three-qubit system. 

The relevant results obtained for $n=3$ to $n=6$ are placed in the Table. I.  
\begin{table}[ht]
\centering
\begin{tabular}{|c|c|c|c|c|}
\hline\hline
n-value& PNC bound &  for qubit  & for 2-qubit & for 3-qubit\\
\hline
\hline
2 & 2 & 2$\sqrt{2}$ & 2$\sqrt{2}$ & 2$\sqrt{2}$ \\
\hline
3 & 4 & 4$\sqrt{3}$ & 4$\sqrt{3}$ & 4$\sqrt{3}$ \\
\hline
4 & 8 & 15.45 & 16 & 16 \\
\hline
5 & 16 & 34.17 & 35.77 & 35.77 \\
\hline
6 & 32 & 71.79 & 77.25 & 78.11 \\
\hline\hline
\end{tabular}
\caption{ The maximum quantum values of the Bell expressions for $n=2$ to $n=6$ are provided for qubit, two-qubit and three-qubit systems. Here PNC in second column denotes the preparation non-contextuality. It is shown that the Bell expressions for $n=4,5$ certify the qubit and two-qubit systems, and the Bell expression for $n=6$ certifies the qubit, two-qubit and three-qubit systems.}
\end{table}

\section{Sharing of preparation contextuality by multiple Bobs}
Let us now examine how many Bobs can sequentially share preparation contextuality if the dimension of the system is lower than the dimension required in obtaining the optimal quantum value  $(\mathcal{B}_{n})_{Q}$.  The notion of sharing of non-local quantum correlation by multiple Bobs was recently put forwarded \cite{silva} where an entangled pair of particles is shared between a single Alice and multiple Bobs performs unsharp measurement on the same particle sequentially. Sharing the non-local correlation by larger number of sequential Bobs requires their sequential measurements to be as unsharp as possible but just enough for violating the preparation non-contextual bound of the family of Bell expressions. In \cite{silva}, it is demonstrated that non-locality through the violation CHSH inequality can be shared by at most two Bobs for the unbiased choices of measurement settings and experimental verification is also reported \cite{schiavon,hu}.  This initiated the studies of sharing of entanglement \cite{bera} and steering \cite{sasmal}. For a suitable choice of entangled state in higher dimension, steering can be shared by unbounded number of Bobs \cite{shenoy} .  One of the authors has earlier demonstrated \cite{asmita19} the sharing of preparation contextuality by arbitrary number of Bobs by using the family of Bell expressions given by Eq. (\ref{nbell}). However, $2^{\lfloor n/2\rfloor}$ dimensional system was taken for optimizing the Bell expression in Eq. (\ref{nbell}).

A relevant question could be to examine the sharing of preparation contextuality by considering the dimension of the system lower than the dimension $d=2^{\lfloor n/2\rfloor}$. Here, we demonstrate that the sharing of preparation contextulity using the Bell expressions  in Eq. (\ref{nbell}) for $n=4,5$ and $6$ for the system having dimensions $d=2^{m}$ where $m=1,2... \lfloor n/2\rfloor$.  
\begin{figure}[ht]
 \centering 
 \includegraphics[width=1.0\linewidth]{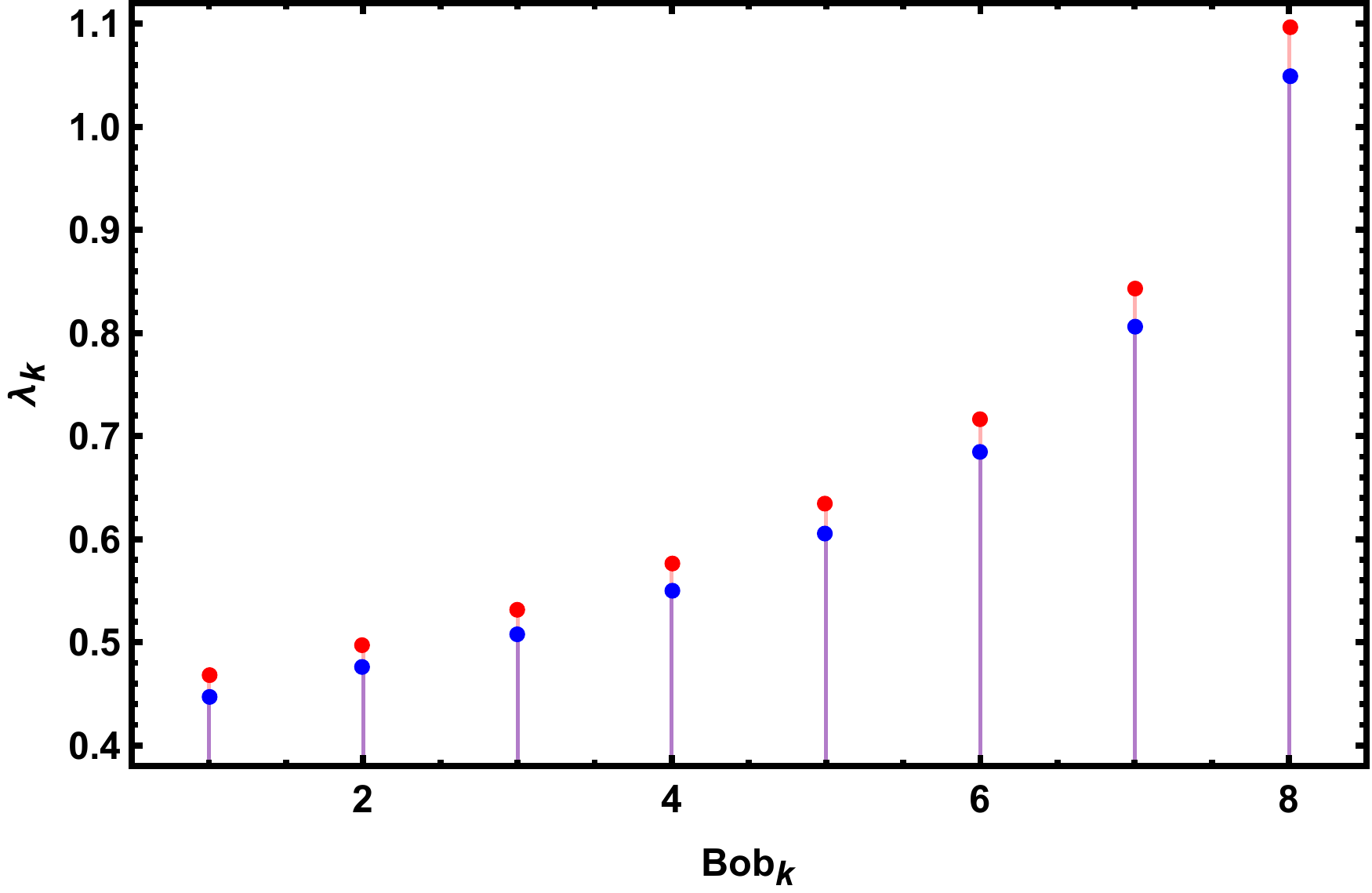}
   \caption{(Color-online) Critical values of unsharpmness parameter $\lambda_{k}$  required for violating the preparation non-contextual bound are plotted $k^{th}$ Bob in the case of $n=5$. Here, blue and red dots denote the critical values corresponding to two-qubit and qubit system respectively. }
 \end{figure}
\begin{figure}[ht]
 \centering 
 \includegraphics[width=1.0\linewidth]{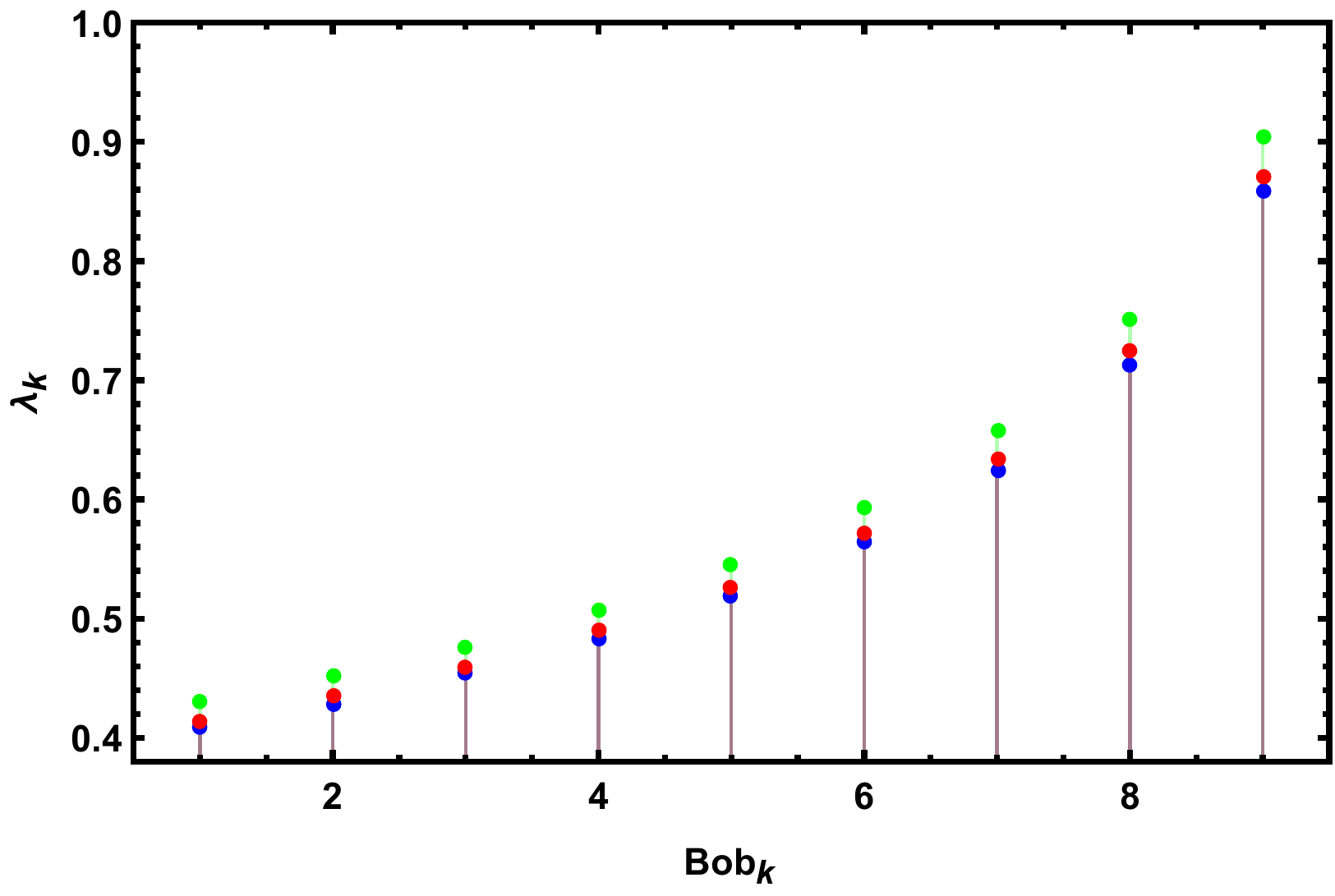}
   \caption{(Color-online) Critical values of unsharpmness parameter $\lambda_{k}$  required for violating the preparation non-contextual bound are plotted $k^{th}$ Bob in the case of $n=6$. Here, blue, red and green dots denote the critical values corresponding to three-qubit, two-qubit and qubit system respectively.}
 \end{figure}
 
  In order to find the number of independent sequential Bobs ($k$) who can share  the preparation contextuality, let us consider that there is one Alice who performs sharp measurement and $k$ number of Bobs perform unsharp measurement sequentially. However, the $k^{th}$ Bob may perform a projective measurement.  For the Bell expression $\mathcal{B}_{n}$ Alice and each Bob  perform the measurements $2^{n-1}$ and $n$ number of dichotomic observable respectively. Given a $n$ value, each Bob requires to perform same set of $n$ number of observables. We also consider that Bob's choices of measurement settings are completely random. Considering Alice and first Bob (say, $Bob_{1}$) share a maximally entangled state and $k-1$ number of Bobs perform the unsharp measurements of the observables $B_{n,y}$ are given by
 \begin{equation}
E_{B_{n},y,j}^{\pm}  = \dfrac{1\pm\lambda_{n,j}}{2} M_{B_{n},y}^{0} + \dfrac{1\mp\lambda_{n,j}}{2} M_{B_{n},y}^{1}
 \end{equation}  
where, $E_{B_{n},y,j}^{\pm}$ are unbiased POVMs and $M_{B_{n},y}^{0}$, $M_{B_{n},y}^{1}$ are the projectors of Bob's observable. Here, $\lambda_{n,j}\in [0,1]$ is the unsharpness parameter for $j^{}$th Bob where $j=1,2.....k-1$ \cite{busch, kumari19}. We consider that for a given $n$, the unsharpness parameter is same for each of Bob's observables $B_{n,y}$  and independent of $y$. 

The shared state between Alice and ${k}^{th}$ Bob is obtained after the unsharp  measurements of $k-1$ Bobs is given by
\begin{eqnarray}
\nonumber
\rho_{n,k}  &=& \frac{1}{n}\sum_{b \in \left\{ + ,- \right\}}\sum_{y = 1}^{n} (\mathbb{I} \otimes \sqrt{ {E}^{b}_{B_{n,y,k-1}}}) \rho_{n,k-1} (\mathbb{I} \otimes \sqrt{ {E}^{b}_{B_{n,y,k-1}}})  \\ \nonumber 
&=& \sqrt{1-\lambda_{n,k-1}^2}\rho_{n,k-1}+\frac{(1- \sqrt{1-\lambda_{n,k-1}^2})}{n}\times\\
&&\sum_{b \in \left\{ + ,- \right\}}\sum_{y = 1}^{n} (\mathbb{I} \otimes {\Pi}^{b}_{B_{n,y,k-1}}) \rho_{n,k-1} (\mathbb{I} \otimes  {\Pi}^{b}_{B_{n,y,k-1}})
\end{eqnarray}
where $\rho_{n,k-1}$ is the state shared between Alice and ${(k-1)}^{th}$ Bob before $(k-1)^{th}$ Bob's unsharp measurement. For for $k^{th}$ sequential Bob the maximum quantum value of the Bell expression given by Eq. (\ref{nbell}) for $d=2^{m}$ dimensional system   can be written as 
\begin{equation}
(\mathcal{B}_{n}^{k})_{Q}^{\lambda} =(\mathcal{B}_{n})_{d=2^{m}}^{max}  \left( \prod_{j=1}^{k-1}  (1+(n-1)\sqrt{1-\lambda_{n,j}^{2}})\right)\lambda_{n,k} 
\end{equation}
where $m=1,2... 2^{\lfloor n/2\rfloor}$. Now, by considering the preparation non-contextual bound $(\mathcal{B}_{n})_{pnc}=2^{n-1}$, the condition on unsharpness parameter for sharing the preparation contxtuality by $k^{}$th Bob is given by 
\begin{equation}
\lambda_{n,k} > \frac{2^{n-1}}{(\mathcal{B}_{n})_{d=2^{m}}^{max}  \left( \prod_{j=1}^{k-1}  (1+(n-1)\sqrt{1-\lambda_{n,j}^{2}})\right) }
\end{equation}
 In order to find how many Bobs can sequentially share preparation contextuality for a given dimension $m=1,2... 2^{\lfloor n/2\rfloor}$, we just need to find the values of $\lambda_{n,j}$ within its valid range $[0,1]$. For this, one needs to use the critical value of the $\lambda_{n,j}$ for $j^{th}$ Bob so that it is just enough to violate the preparation contextuality. 

For $n=5$, if dimension of Hilbert space for local system is $d=2$, i.e., qubit system, we find that the sharing of preparation contextuality is possible for at most seven Bob (Figure. 1). Importantly, instead of qubit system, if two-qubit system is taken, the number of Bobs sharing preparation contextuality remains same. Note that, Bell expression for $n=5$ reaches its optimal value for the two-qubit system. However, for every $j$, the value of unsharpness parameter required for sequential violation of preparation non-contextual bound is larger for qubit system, as shown in Figure 1.  Similar feature is also obtained from the Bell expression for $n=6$. It can be seen from the Figure 2, that equal number of Bobs can sequentially share the preparation contextuality for qubit, two-qubit and three-qubit systems. We may conjecture that this feature remains same for any arbitrary $n$. For this, one needs to find the maximum quantum value of $(\mathcal{B}_{n})_{Q}$  for $d=2^{m}$ dimensional systems where $m=1,2... \lfloor n/2\rfloor$. Thus, sharing of preparation contextuality using qubit system is advantageous in the sense that one requires to deal with a lower dimensional system.

\section{Summary and discussions}

In summary, we have provided a family of Bell expressions that can certify various  dimensions of quantum system. Such a Bell expressions were derived based on a two-party communication game known as $n$-bit parity-oblivious random access code \cite{spek09}. It can be shown that the success probability of that game can be solely determined by the aforementioned family of Bell expressions $(\mathcal{B}_{n})$ where Alice and Bob use $2^{n-1}$ and $n$ number of dichotomic measurements respectively. In a RAC, the parity oblivious condition  implies that Alice may communicate any number ($<n$) of bit  but such a communication does not allow Bob to retrieve the information about the parity of the Alice's inputs. It is shown \cite{spek09} that the parity oblivious constraint is \cite{spekk05} equivalent to  preparation non-contextuality assumption in an ontological model. We have shown that for a given $n$, such a constraint on the encoding scheme  reduces the local bound of the family of Bell expressions $\mathcal{B}_{n}$ to preparation non-contextual bound \cite{spekk05}.  For a given $n$, the optimal quantum value of the Bell expression $\mathcal{B}_{n}$ can only be achieved for the quantum system having local Hilbert space dimension $d=2^{\lfloor n/2\rfloor}$. 

Note that, the Bell expressions for $n=2$ and $n=3$ reduce to CHSH \cite{chsh} and elegant Bell expressions \cite{gisin} respectively, which cannot serve as dimension witness as they can be optimized for qubit system. However, each of the Bell inequalities for $n\geq 4$ distinguish  the dimensions $d=2^{m}$ of the Hilbert space with $m=1,2..2^{\lfloor n/2\rfloor}$.  We provided explicit examples by considering the Bell expressions for $n=4,5$ and $6$. It is shown that for both $n=4$ and $n=5$ the respective Bell expressions $\mathcal{B}_{4}$ and $\mathcal{B}_{5}$ certify qubit and two-qubit systems. But, for $n=6$, we have found that the relevant Bell expressions $\mathcal{B}_{6}$ certifies the qubit, two-qubit and three-qubit local systems.  

Further, we have examined the sharing of preparation contextuality by multiple sequential Bobs through the violation of $(\mathcal{B}_{n})_{pnc}$ in Eq. (\ref{pncbound})  when the dimension of the system is lower than that is required for achieving the optimal quantum value   $(\mathcal{B}_{n})_{Q}^{opt}$. One of us have shown \cite{asmita19} that the sharing of preparation contextuality can be demonstrated by arbitrary number of Bobs by using optimal quantum value  $(\mathcal{B}_{n})_{Q}^{opt}$ for $d=2^{\lfloor n/2\rfloor}$ dimensional Hilbert space. Here, by providing the examples of $n=5$ and $n=6$, we have demonstrated that even for lower dimensional system the number of Bobs who can share the preparation contextuality remains same but the value of unshrapness parameter required is always higher in lower dimensional system.  

Finally, we remark that from our study it is straightforward to understand that for any arbitrary $n$, the family of Bell expressions  $\mathcal{B}_{n}$ can certify the Hilbert space having dimension $2^{m}$ with $m=1,2... 2^{\lfloor n/2\rfloor}$. For this, following the scheme presented here the maximum quantum value  $(\mathcal{B}_{n})_{d=2^{m}}$ for different $m$ value has to be derived. The analytical derivation can be lengthy with increasing value of $n$ but it is doable to some extend. The numerical technique can also be an obvious option. It would then be interesting to examine if the sharing of preparation contexuality can be demonstrated through the Bell expression $\mathcal{B}_{n}$ by unbounded number of Bobs even for a qubit system. This would spur the experimental test of the sharing of preparation contextuality for unbounded number of sequential observers. Studies along this line could be an interesting avenue for further research.

\section*{Acknowledgments}
AKP acknowledge the support from the project DST/ICPS/QuST/Theme 1/2019/4. 
SSM acknowledges the support from Ramanujan Fellowship research grant SB/S2/RJN-083/2014. 
\begin{widetext}
\appendix
\section{Explicit parity-oblivious conditions for $n=4$}
Here we provide the details of the calculation how the choices of Alice's observables that provides the maximum quantum value of $\mathcal{B}_{4,y}$ for qubit and two-qubit system satisfy the parity-oblivious condition given by Eq. (\ref{oba}). The Alice's choices of observables can be found by using Eq. (\ref{conda}) given the choices of $B_{4,y}$s. In order to examine this issue, we consider the non-trivial elements of the parity set $s_4 \in \mathbb{P}_{4}$ where $s_{4} =1110,1101,1011$ and $0111$. This dictates that corresponding to each element of $s_{4}$ we have four different functional relations between the Alice's observables satisfying the parity-oblivious condition in Eq. (\ref{oba}), are the following;  
\begin{align}
\label{po}
A_{4,1}+A_{4,2}-A_{4,3}-A_{4,4}-A_{4,5}-A_{4,6}-A_{4,7}+A_{4,8}=0  \nonumber \\
 A_{4,1}-A_{4,2}+A_{4,3}-A_{4,4}-A_{4,5}-A_{4,6}+A_{4,7}-A_{4,8}=0  \nonumber \\
 A_{4,1}-A_{4,2}-A_{4,3}+A_{4,4}-A_{4,5}+A_{4,6}-A_{4,7}-A_{4,8}=0  \nonumber \\
A_{4,1}-A_{4,2}-A_{4,3}-A_{4,4}+A_{4,5}+A_{4,6}+A_{4,7}+A_{4,8}=0  
\end{align} 
For a particular case say, $s=1110$, using Eq. (\ref{cond2}) we find the relations between $\omega_{4,i}$ required to be satisfied are given by

\begin{align}
\label{po1}
\alpha_{4,1}+\alpha_{4,2} -\alpha_{4,3}-\alpha_{4,4}+\alpha_{4,5}-\alpha_{4,6}-\alpha_{4,7}+\alpha_{4,8}=0  \nonumber \\
\alpha_{4,1}+\alpha_{4,2} -\alpha_{4,3}+\alpha_{4,4}-\alpha_{4,5}-\alpha_{4,6}+\alpha_{4,7}-\alpha_{4,8}=0  \nonumber \\
\alpha_{4,1}+\alpha_{4,2} +\alpha_{4,3}-\alpha_{4,4}-\alpha_{4,5}+\alpha_{4,6}-\alpha_{4,7}-\alpha_{4,8}=0  \nonumber \\
\alpha_{4,1}-\alpha_{4,2} -\alpha_{4,3}-\alpha_{4,4}-\alpha_{4,5}+\alpha_{4,6}+\alpha_{4,7}+\alpha_{4,8}=0  
\end{align} 
Here $\alpha_{4,i}=1/\omega_{4,i}$ Similar set of four relations between $\alpha_{4,i}$ can be found for each of the other elements of $s_{4}$. There are two conditions on $\omega_{4,i}$s are available for which four equations in Eq. (\ref{po1}) will be simultaneously  satisfied. First,  if $\omega_{4,1}=\omega_{4,2}=\omega_{4,5}=\omega_{4,6}=\omega_{4}^{\prime}$ and $\omega_{4,2}=\omega_{4,3}=\omega_{4,7}=\omega_{4,8}=\omega_{4}^{\prime\prime}$, and second  if for every $i$ the $\omega_{4,i}$ is the same. Note that, second condition requires four mutually anti-commuting observables and thus cannot be satisfied for qubit system. Similar relations can be obtained for other three elements of  $s_{4}$. Using the first restriction on $\omega_{4,i}$, we obtain Eq. (\ref{rac4}) and this in turn proves that the parity-oblivious condition is satisfied by Alice's choices of observables.
\section{Detailed calculation for $n=5$}
For $n=5$ the Bell expression  involves the measurements of eight and five dichotomic observables by Alice and Bob respectively. From Eq. (\ref{nbell}), the Bell expression  can be writen as 

\ba
\label{b5}
	\mathcal{B}_{5}&=&A_1\otimes ( B_1 +B_2 +B_3 +B_4+B_5)+ A_2 \otimes ( B_1 +B_2 +B_3 +B_4-B_5)+A_3 \otimes (B_1 +B_2 +B_3 -B_4+B_5)\nonumber \\
	&+&A_4\otimes ( B_1 +B_2 +B_3 -B_4-B_5)+A_5\otimes ( B_1 +B_2 -B_3 +B_4+B_5) +A_6 \otimes( B_1 +B_2 -B_3 +B_4-B_5)\nonumber \\
	&+&A_7\otimes( B_1 +B_2 +B_3 -B_4-B_5)+A_8\otimes( B_1 +B_2 -B_3 -B_4+B_5)+A_9 \otimes( B_1 -B_2 -B_3 +B_4+B_5)\nonumber \\
	&+&A_{10} \otimes(-B_1 -B_2 +B_3 +B_4+B_5) + A_{11} \otimes( B_1 +B_2 -B_3 +B_4-B_5)+ A_{12} \otimes(B_1 -B_2 +B_3 -B_4+B_5)\nonumber \\
	&+&A_{13} \otimes(B_1 -B_2 +B_3 +B_4-B_5) +A_{14}\otimes(-B_1 +B_2 -B_3 +B_4+B_5)+ A_{15} \otimes(- B_1 +B_2 +B_3 +B_4-B_5)\nonumber \\
	&+& A_{16}\otimes(-B_1 +B_2 +B_3 -B_4+B_5)
	\ea
whose preparation non-contextual bound is $	(\mathcal{B}_{5})_{pnc}\leq 16$. As mentioned in the main text that Alice's observables need to satisfy the parity oblivious conditions given by Eq. (\ref{cond2}). The parity set $\mathbb{P}_{5}$ contains $11$ non-trivial elements and each of them provides a functional relationship between the Alice's choice of observables $A_{5,i}$. For example, if we take one of the elements, say $11111$, a functional relation between  $A_{n,i}$s has to be satisfied by Alice's observables is given by
	\ba
	\label{s11111}
	A_{4,1}-A_{4,2}-A_{4,3}-A_{4,4}-A_{4,5}-A_{4,6}+\sum\limits_{i=7}^{16} A_{n,i}=0
	\ea
	Similar ten more such constraints can be found for other elements $s_5 \in \mathbb{P}_n$. Note that the condition for optimization required for SOS approach is given by
	\ba
\label{condap}
	\forall i \  \  \  \sum\limits_{y=1}^{n} (-1)^{x^i_y} \alpha_{n,i} B_{n,y} |\psi\rangle =A_{n,i}|\psi\rangle
\ea
where $\alpha_{5,i}=1/\omega_{5,i}$. In the present case of $n=5$, by using Eq. (\ref{condap}), from Eq. (\ref{s11111}) we have following conditions on $\alpha_{5,i}$s that  need to be satisfied are given by

\ba
\label{omega51}
&&\alpha_{5,1}-\alpha_{5,2}-\alpha_{5,3}-\alpha_{5,4}-\alpha_{5,5}+\alpha_{5,6}+\alpha_{5,7}\\
\nonumber
&+&\alpha_{5,8}+\alpha_{5,9}-\alpha_{5,10}+\alpha_{5,11}+\alpha_{5,12}+\alpha_{5,13}-\alpha_{5,14}-\alpha_{5,15}-\alpha_{5,16}=0\\
\label{omega52}
&&\alpha_{5,1}-\alpha_{5,2}-\alpha_{5,3}-\alpha_{5,4}+\alpha_{5,5}-\alpha_{5,6}+\alpha_{5,7}+\alpha_{5,8}
\\
\nonumber
&-&\alpha_{5,9}-\alpha_{5,10}+\alpha_{5,11}-\alpha_{5,12}-\alpha_{5,13}+\alpha_{5,14}+\alpha_{5,15}+\alpha_{5,16}=0\\
\label{omega53}
&&\alpha_{5,1}-\alpha_{5,2}-\alpha_{5,3}+\alpha_{5,4}-\alpha_{5,5}-\alpha_{5,6}+\alpha_{5,7}-\alpha_{5,8}
\\
\nonumber
&-&\alpha_{5,9}+\alpha_{5,10}-\alpha_{5,11}+\alpha_{5,12}+\alpha_{5,13}-\alpha_{5,14}+\alpha_{5,15}+\alpha_{5,16}=0\\
\label{omega54}
&&\alpha_{5,1}-\alpha_{5,2}+\alpha_{5,3}-\alpha_{5,4}-\alpha_{5,5}-\alpha_{5,6}-\alpha_{5,7}-\alpha_{5,8}
\\
\nonumber
&+&\alpha_{5,9}+\alpha_{5,10}+\alpha_{5,11}-\alpha_{5,12}+\alpha_{5,13}+\alpha_{5,14}+\alpha_{5,15}-\alpha_{5,16}=0\\
\label{omega55}
&&\alpha_{5,1}+\alpha_{5,2}-\alpha_{5,3}-\alpha_{5,4}-\alpha_{5,5}-\alpha_{5,6}-\alpha_{5,7}+\alpha_{5,8}
\\
\nonumber
&+&\alpha_{5,9}+\alpha_{5,10}-\alpha_{5,11}+\alpha_{5,12}-\alpha_{5,13}+\alpha_{5,14}-\alpha_{5,15}+\alpha_{5,16}=0
\ea

The functional relations between $\omega_{5,i}$ given by Eqs. (\ref{omega51}-\ref{omega55}) provide two solutions; first, $\omega_{5,1}=\omega_{5,2}=\omega_{5,4}=\omega_{5,5} =\omega_{5,9} =\omega_{5,11} =\omega_{5,12} =\omega_{5,15}=\omega_{5}^{\prime}$ and $\omega_{5,3}=\omega_{5,6}=\omega_{5,7}=\omega_{5,8}=\omega_{5,10} =\omega_{5,13} =\omega_{5,14} =\omega_{5,16}=\omega_{5}^{\prime\prime}$ and second one is that  $\omega_{5,i}$ is equal to each other for every $i$. Note that the second condition cannot be satisfied by the observables in a  qubit system. Using the first solution, we find the following relations between $B_{5,y}$ needs to be satisfied; $\{B_{5,1},B_{5,2}\}=\{B_{5,2},B_{5,3}\}=\{B_{5,1},B_{5,3}\}=\{B_{5,3},B_{5,4}\}=\{B_{5,3},B_{5,5}\}=\{B_{5,4},B_{5,5}\}=0$, $\{B_{5,1},B_{5,4}\}=\{B_{5,1},B_{5,5}\}$ and $\{B_{5,2},B_{5,4}\}=-\{B_{5,2},B_{5,5}\}$. Using those relations and Eq. (\ref{optbn}), we can write

\ba
\label{rac6}
(\mathcal{B}_{5})_{Q}\leq 4\left(\omega_{5}^{\prime}+\omega_{5}^{\prime\prime}\right)=\Big( \sqrt{5+ \{\left(B_{5,1}+B_{5,2}\right),\left(B_{5,4}+B_{5,5}\right)\}} +\sqrt{5+\{\left(B_{5,1}-B_{5,2}\right),\left(B_{5,4}-B_{5,5}\right)\}}\Big)
\ea 
which is Eq. (\ref{rac5}) in the main text. Thus, the Alice's observables maximizing $(\mathcal{B}_{5})_{Q}$ satisfy the parity-oblivious condition.
	\end{widetext}

\end{document}